\documentclass[%
reprint,
superscriptaddress,
showpacs,preprintnumbers,
 amsmath,amssymb,
 aps,
prb,
]{revtex4-1}
\usepackage{graphicx}
\usepackage{dcolumn}
\usepackage{bm}
\usepackage{color}

\newcommand{\diff}{\ensuremath{\text{d}}}

\newcommand{\ket}[1]{\ensuremath{|#1\rangle }}
\newcommand{\hpr}[2]{\ensuremath{\langle #1|#2\rangle}}

\begin{document}


\title{Multi-scale magnetic study on Ni(111) and graphene on Ni(111)}

\author{L.~V.~Dzemiantsova}
 \email{email address: ldzemian@physnet.uni-hamburg.de}
  \affiliation{%
 Institute of Applied Physics, University of Hamburg, Jungiusstrasse 11, D-20355 Hamburg, Germany
}%
\author{M.~Karolak}
 \email{email address: mkarolak@physik.uni-hamburg.de}
 \affiliation{%
 I.~Institute of Theoretical Physics, University of Hamburg, Jungiusstrasse 9, D-20355 Hamburg, Germany
}%
\author{F.~Lofink}
 \email{email address: flofink@physnet.uni-hamburg.de}
\author{A.~Kubetzka}
 \affiliation{%
 Institute of Applied Physics, University of Hamburg, Jungiusstrasse 11, D-20355 Hamburg, Germany
}%
\author{B.~Sachs}
 \affiliation{%
 I.~Institute of Theoretical Physics, University of Hamburg, Jungiusstrasse 9, D-20355 Hamburg, Germany
}%
\author{K.~von~Bergmann}
 \affiliation{%
 Institute of Applied Physics, University of Hamburg, Jungiusstrasse 11, D-20355 Hamburg, Germany
}%
\author{S.~Hankemeier}
 \affiliation{%
 Institute of Applied Physics, University of Hamburg, Jungiusstrasse 11, D-20355 Hamburg, Germany
}%
\author{T.~O.~Wehling}
 \affiliation{%
 I.~Institute of Theoretical Physics, University of Hamburg, Jungiusstrasse 9, D-20355 Hamburg, Germany
}%
\author{R.~Fr\"{o}mter}
 \affiliation{%
 Institute of Applied Physics, University of Hamburg, Jungiusstrasse 11, D-20355 Hamburg, Germany
}%
\author{H.\,P.~Oepen}	
 \affiliation{%
 Institute of Applied Physics, University of Hamburg, Jungiusstrasse 11, D-20355 Hamburg, Germany
}%
\author{A.\,I.~Lichtenstein}
 \affiliation{%
 I.~Institute of Theoretical Physics, University of Hamburg, Jungiusstrasse 9, D-20355 Hamburg, Germany
}%
\author{R.~Wiesendanger}	
 \affiliation{%
 Institute of Applied Physics, University of Hamburg, Jungiusstrasse 11, D-20355 Hamburg, Germany
}%

\date{\today}

\begin{abstract}
We have investigated the magnetism of the bare and graphene-covered (111) surface of a Ni single crystal employing three different magnetic imaging techniques and \emph{ab initio} calculations, covering length scales from the nanometer regime up to several millimeters. With low temperature spin-polarized scanning tunneling microscopy (SP-STM) we find domain walls with widths of 60~-~90~nm, which can be moved by small perpendicular magnetic fields. Spin contrast is also achieved on the graphene-covered surface, which means that the electron density in the vacuum above graphene is substantially spin-polarized. In accordance with our \emph{ab initio} calculations we find an enhanced atomic corrugation with respect to the bare surface, due to the presence of the carbon $p_z$ orbitals and as a result of the quenching of Ni surface states. The latter also leads to an inversion of spin-polarization with respect to the pristine surface. Room temperature Kerr microscopy shows a stripe like domain pattern with stripe widths of 3~-~6~$\mu$m. Applying in-plane-fields, domain walls start to move at about 13~mT and a single domain state is achieved at 140~mT. Via scanning electron microscopy with polarization analysis (SEMPA) a second type of modulation within the stripes is found and identified as 330~nm wide V-lines. Qualitatively, the observed surface domain pattern originates from bulk domains and their quasi-domain branching is driven by stray field reduction.
\end{abstract}


\maketitle

\section{\label{sec:intro}Introduction}
The knowledge of the magnetic properties of 3d materials is crucial for the design of micromagnetic devices and the tailoring of their properties. In particular, the interest in Ni(111) was recently renewed from both the experimental and theoretical sides. The Shockley type spin-split surface state of Ni(111) was reported to play an important role for the magnetic properties of the surface.\cite {okuda_2009,nishimura_2009,braun_2008,braun_2002} However, even though bulk Ni(111) is a conventional ferromagnet, its domain structure is not well investigated. Since there are no easy magnetization axes in the plane, a complex magnetic pattern is expected. \cite{getzlaff_2008}  A model of a multiple quasi-domain branching was proposed for a Ni(111) platelet by Hubert and Sch\"afer.\cite{hubert_1998} For a long time, only bitter technique data has been available, which gives only a rough idea of the surface domain structure.\cite{elschner_1955} Ni(111) was more recently studied by Magnetic Force Microscopy (MFM), where domains of the order of 500 nm were observed at $T=8$~K and domain walls shifted in perpendicular fields of $B=25$ mT.\cite{kaiser_2007} Still, a complete picture of the microscopic domain pattern is missing.

Ni(111) has become again the focus of current research due to its role as a perfect substrate for the growth of graphene.\cite{backer_1995} Owing to the very small lattice mismatch, it grows pseudomorphically, and it has been recently shown that, due to the strong hybridization with the Ni atoms, graphene-covered Ni can be an efficient source of spin-polarized electrons.\cite{dedkov_rudiger_2008,dedkov_2008} In addition, it was also reported that graphene passivates Ni(111) against oxygen exposure,\cite{dedkov_2008} which makes the graphene/Ni(111) system a promising candidate for applications in carbon-based magnetic media and spintronic devices.

In this study, we employ three different magnetic imaging techniques to investigate Ni(111) and graphene on Ni(111), covering length scales from the nanometer regime up to several millimeters. With scanning tunneling microscopy (STM) the atomic lattice can be resolved and bare Ni(111) is characterized by scanning tunneling spectroscopy (STS) where occupied and unoccupied local density of states (LDOS) are probed. Spin-polarized STM (SP-STM)\cite{wiesendanger_2009} can access single domain walls and their response to an applied magnetic field. To investigate the entire surface domain structure, however, techniques with a larger field of view are necessary. We used scanning electron microscopy with polarization analysis (spin-SEM or SEMPA)\cite{koike_1984} for vectorial mapping of the surface magnetization and Kerr microscopy for field dependent imaging.\cite{hubert_1998} Interestingly, SP-STM and SEMPA can access the Ni magnetism through the graphene layer, and we show that it is unchanged compared to bare Ni(111). For a deeper understanding of the two surfaces we performed spin-resolved density functional theory (DFT) calculations. Evaluating the calculated LDOS in the vacuum above the surfaces allows the interpretation of the STM and STS data. As a result of surface state quenching, we predict an inversion of the spin-polarization above the graphene layer with respect to the pristine Ni(111) surface.

\section{\label{sec:expdetails}Experimental details}
The same Ni(111) single crystal ($3~\textnormal{mm}\times7~\textnormal{mm}$ width, 1 mm thickness) was used in all experiments. It was cleaned by repeated cycles of 800 eV Ar$^+$ ion etching at room temperature (RT) and annealing at $T=1100$~K. The sample was considered clean when no impurities such as carbon or sulfur were detected by Auger electron spectroscopy (AES) and a hexagonal ($1\times1$) pattern was observed by low-energy electron diffraction (LEED). The graphene layer was grown on Ni(111) by chemical vapor deposition (CVD):\cite{grunes_2009} Ni(111) was heated to $T=950$~K in an ethylene atmosphere (C$_2$H$_4$) of $p=5\times10^{-7}$~mbar for 200 s (100 L) and subsequently allowed to cool in ultra-high vacuum (UHV).

The STM experiments were carried out in a multi-chamber UHV system with separate chambers for ion etching, CVD graphene growth, and STM measurements. One microscope operates at RT\cite{witt_1997} while another is thermally connected to a liquid-He bath cryostat reaching $T=8.0 \pm0.5$~K and features an out-of-plane magnetic field of up to $B=\pm 2.5$~T.\cite{pietzsch_2000} It is equipped with a tip exchange mechanism and for spin-averaged measurements we used W tips, that were flashed \emph{in vacuo} to approx.~$T=2300$~K. For our SP-STM studies these tips were then coated with about 50 atomic layers (AL) of Cr, and annealed at $T\approx500$~K for 5 min. An antiferromagnetic tip coating is chosen to minimize the magnetostatic interaction between the probe and the magnetic sample. Constant-current ($I$) images (topography) and maps of differential conductance ($\diff I / \diff U$)~\cite{tersoff_1983,li_1997} were measured simultaneously with closed feedback loop using lock-in technique by adding a modulation voltage ${U}_\textnormal{mod}=25$~mV to the sample bias $U$. Single $\diff I / \diff U$ spectra were taken at specific positions with the tip-sample distance stabilized at ${U}_\textnormal{stab}$ and ${I}_\textnormal{stab}$ before switching off the feedback loop.

SEMPA was used for simultaneous vectorial mapping of both orthogonal in-plane magnetization components at the surface.\cite{froemter_2011} Magnetization images of 20 nm lateral resolution were acquired at RT using a primary beam of 6 nA at 8 keV. The SEMPA instrument is located in a separate UHV-facility at $5\times10^{-11}$~mbar base pressure, into which graphene-coated Ni(111) samples were transferred through air. To remove the graphene layer, Ar$^+$ ion etching at 600 eV without post annealing was used. For contrast enhancement 4 AL of Fe were deposited from an e-beam evaporator at 0.2 AL/min. 

We used a full-field Kerr microscope with white-light illumination \cite{evico} working at ambient conditions to investigate large areas up to several millimeters on the Ni surface. Kerr microscopy utilizes the magneto-optic Kerr effect (MOKE) to visualize the magnetic surface structure of an investigated sample. An arbitrary magnetization component can be selected for imaging by using appropriate apertures in the back focal plane. A lateral resolution of 300 nm was achieved and imaging in external fields was possible.

\section{\label{sec:stm}STM results}
\subsection{\label{sec:ni111}Ni(111)}
\begin{figure}[t]
\includegraphics[width=1\columnwidth]{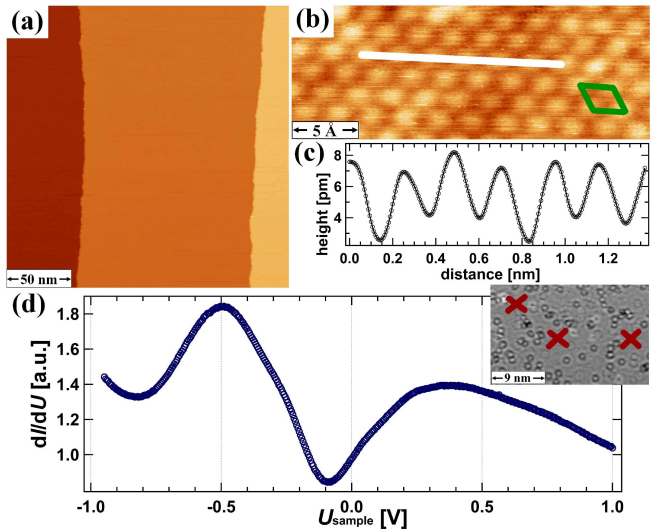}
\caption{\label{fig:pic_1} {(color online) Ni(111), spin-averaged data: (a)~Constant-current image ($U=-1$~V, $I=0.5$~nA) taken at $T=8$~K. (b) Atomically resolved image ($U=-4$~mV, $I=5$~nA) taken at RT. The diamond highlights the unit cell. (c)~Height profile along the line displayed in (b). (d)~The $\diff I / \diff U$ spectrum (${U}_\textnormal{stab}=+1$~V, ${I}_\textnormal{stab}=2$~nA, ${U}_\textnormal{mod}=80$~mV) is averaged over data taken at three different locations marked as red crosses in the $\diff I / \diff U$ map (inset). }}
\end{figure}

Figure~\ref{fig:pic_1}(a)~shows a typical surface area of bare Ni(111) including two monatomic steps. Terrace widths vary with lateral position in the range of 20~-~200~nm.  As seen in Fig.~\ref{fig:pic_1}(b) the atomic lattice can be resolved on the terraces. Fig.~\ref{fig:pic_1}(c)~displays a line profile along a closed packed row (white line in (b)) and the interatomic spacing is in agreement with the nearest-neighbor atomic distance of Ni (2.49~\AA). As expected for closed packed surfaces like fcc(111), the corrugation  is comparatively small and lies in the range of 3~-~5~pm.\cite{wintterlin_1989}

Despite the nice ordering of surface Ni atoms, residual contamination is present in the sample, in particular sub-surface defects. These defects have very low corrugation in constant current images, but are clearly seen in $\diff I / \diff U$ maps, e.g.~at $U=+1$~V (see the inset in Fig.~\,\ref{fig:pic_1}(d)). We therefore took care to measure $\diff I / \diff U$ spectra on defect free areas: Fig.~\,\ref{fig:pic_1}(d) shows an average of three spectra measured at positions as indicated in the inset. We observe two broad maxima, 500~meV below and about 400~meV above the Fermi level, $E_\textnormal{F}$, respectively. The spectrum agrees well with the experimental results of K.-F.~Braun and co-workers.\cite{braun_2008} Our first-principle calculations (Section \ref{sec:theory}) attribute these features to a minority spin surface resonance below and the Shockley state of both spin character above $E_\textnormal{F}$.

\begin{figure}[t]
\includegraphics[width=1\columnwidth]{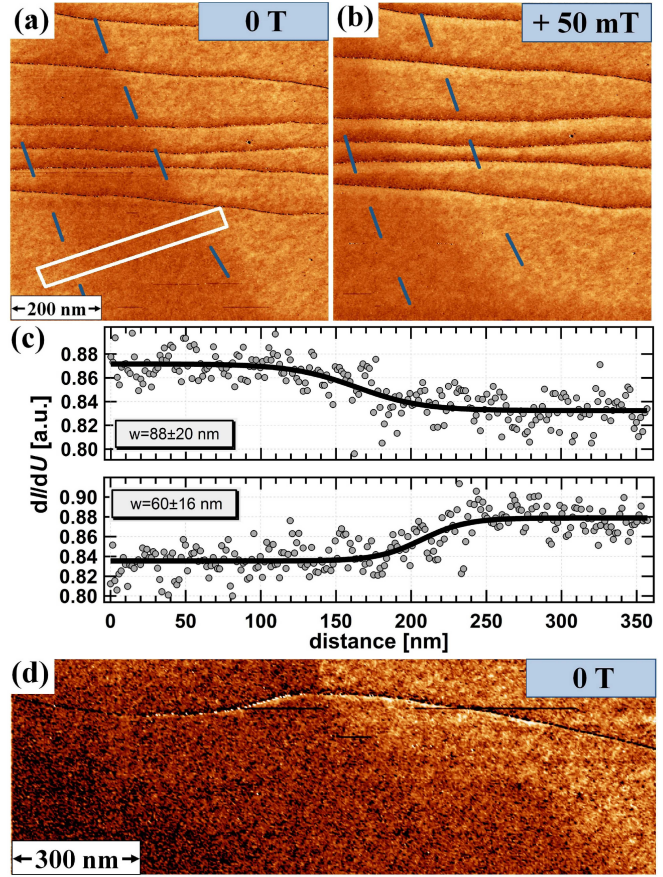}
\caption{\label{fig:pic_2} {(color online) Ni(111), spin-polarized data: Magnetic $\diff I / \diff U$ maps taken at (a) $B=0$~T and (b) $B=+50$~mT. Dashed lines show the shift of domain walls in the external magnetic field. (c) Line sections across domain walls marked by white box in (a). Gray circles and black lines represent experimental data and fitted profiles, respectively. (d) Magnetic $\diff I / \diff U$ map of 1.5~$\mu$m width taken at $B=0$~T. All data measured at $U=-200$~mV and $I=2.5$~nA.}}
\end{figure}

To investigate the magnetic properties of the Ni(111) surface we use SP-STM: Fig.~\,\ref{fig:pic_2}(a) shows a $\diff I / \diff U$ map of a sample area exhibiting nearly horizontal steps. The darker and brighter regions indicate magnetic domains of Ni(111) and the dashed lines indicate domain walls. To prove the magnetic origin of the observed contrast we apply an out-of-plane magnetic field of $B=50$~mT and as a result both domain walls have moved to the left as seen in Fig.~\ref{fig:pic_2}(b), the right one by about 150 nm. Line sections across the walls, as indicated by a white box in Fig.~\,\ref{fig:pic_2}(a) are shown in Fig.~\,\ref{fig:pic_2}(c). Fitting a standard wall profile\cite{standard_fit} (solid line),
\begin{equation}
\tanh((x)/(w/2)),
\label{eq:one}
\end{equation}
where  $x$ is a lateral distance, to the experimental data (dots) yields wall widths of $w=88\pm 20$~nm and $w=60\pm 16$~nm, respectively. The wall width determined for the wall in (b) at +50 mT is $w=62\pm 17$~nm. This shows that while we can move domain walls in the out-of-plane field of +50 mT the width of the wall in Fig.~\,\ref{fig:pic_2} is not altered within the error of the measured width. The reason why these walls can be moved by the out-of-plane field becomes clear from the volume domain structure as deduced from SEMPA measurements (Section \ref{sec:sempa}). The magnetic contrast, which can be defined as the asymmetry of the $\diff I / \diff U$ signals of bright (b) and dark (d) areas,\cite{wiesendanger_2009}

\begin{equation}
A(U)=\frac{\diff I / \diff U(U)_\textnormal{b}
-\diff I / \diff U(U)_\textnormal{d}}
{\diff I / \diff U(U)_\textnormal{b}
+\diff I / \diff U(U)_\textnormal{d}},
\end{equation}
is typically low on the Ni(111) surface. Evaluating the data in Fig.~\,\ref{fig:pic_2} yields a value of only $A(-200~\textnormal{mV})=2$~\%.

The spin-resolved $\diff I / \diff U$ map in Fig.~\,\ref{fig:pic_2}(d) has a width of 1.5~$\mu$m and shows three areas of different intensity, i.e.~magnetic domains, and an atomic step. The asymmetry between the highest and lowest signal amounts to 2~\%. The occurrence of several domains with different magnetization directions on this length scale indicates already an interesting overall magnetic structure, which requires an imaging technique with a larger field of view.

\subsection{\label{sec:gni111}Graphene on Ni(111)}
\begin{figure}[t]
\includegraphics[width=1\columnwidth]{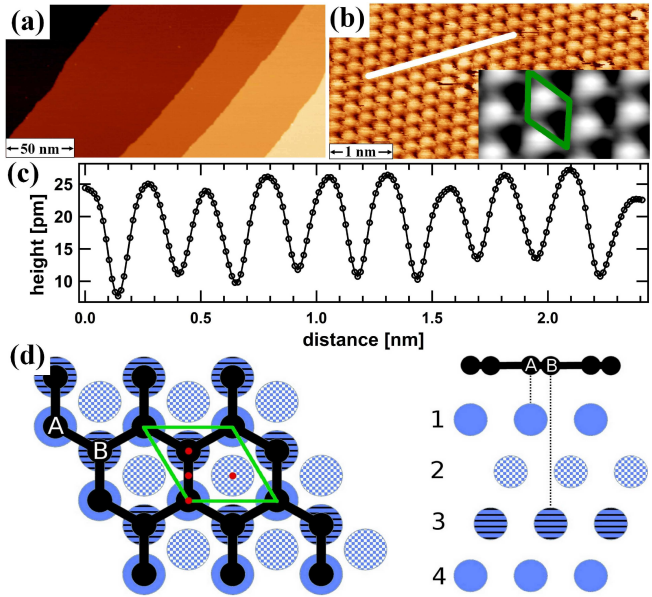}
\caption{\label{fig:pic_3}{(color online) Graphene on Ni(111), spin-averaged data: (a) Constant current overview image ($U=-1$~V, $I=0.5$~nA), taken at $T=8\pm1$~K, showing flat terraces and four monatomic steps. (b) Zoom-in on a terrace ($U=+2.5$~mV, $I=5$~nA) showing the atomic lattice. (c) Height profile along the line depicted in (b). (d) Top and side views of the graphene/Ni top-fcc structure: Black color indicates carbon, blue color Ni atoms. Red dots indicate the positions of the empty spheres (see Section \ref{sec:theory}). The unit cell is highlighted by the diamond.}}
\end{figure}

The graphene layer is commensurate with Ni(111) due to the small lattice mismatch.\cite{bertoni_2004}~A typical sample of graphene on Ni is shown in Fig.~\,\ref{fig:pic_3}(a). The flat terraces indicate perfect single domain graphene formation without any visible defects. At higher magnification in Fig.~\,\ref{fig:pic_3}(b) a triangular lattice is seen rather than the honeycomb structure of the carbon atoms (the unit cell is highlighted by the diamond as shown in the inset). This is not surprising since neighboring carbon atoms occupy non-equivalent sites on the Ni substrate: in the ball model in Fig.~\,\ref{fig:pic_3}(d) the carbon atoms labeled A reside on top of the Ni atoms of the first layer, while the carbon atoms labeled B are at positions of the Ni atoms of the third layer (fcc hollow site).\cite{bertoni_2004,gamo_1997,fuentes_2008}
\begin{figure}[t]
\includegraphics[width=1\columnwidth]{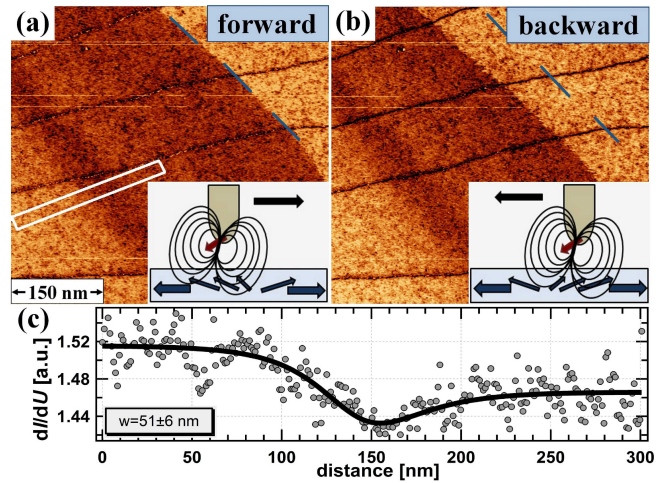}
\caption{\label{fig:pic_4}{(color online) Graphene on Ni(111), spin-polarized data: Magnetic $\diff I / \diff U$ maps measured in (a) forward and (b)~backward scan directions ($U=-200$~mV, $I=0.5$~nA). While the left wall is static, the right one is moved by the stray field of the tip (see text). (c) Line section as indicated in (a) and a fit of a general wall profile,\cite{fit_with_angle,kubetzka_2003} yielding a wall width of about 50 nm. Data taken at $B=0$~T.}}
\end{figure}

The white line in  Fig.~\ref{fig:pic_3}(b) indicates the position of the line profile in  Fig.~\ref{fig:pic_3}(c) and the distance between maxima is the same as for the bare Ni(111) surface shown in Fig.~\,\ref{fig:pic_1}(b,c), reflecting the nearest-neighbor distance of Ni (2.49~\AA). However, in contrast to Fig.~\,\ref{fig:pic_1}(b,c), fcc and hcp hollow sites are now distinguishable, due to the B type atoms on fcc positions. In addition, the atomic corrugation of 10~-~15~pm for graphene on Ni(111) observed here is roughly a factor of 3 larger than that measured on bare Ni(111). Both a triangular lattice structure and an enhanced corrugation seen in STM images are a purely electronic effect originating from graphene $p_z$ states around $E_\textnormal{F}$ and the quenching of Ni surface states as will be discussed in Section \ref{sec:theory}.

To investigate whether we can still measure a magnetic signal on graphene-coated Ni(111) we perform SP-STM measurements. Figures~\ref{fig:pic_4}(a) and (b) show $\diff I / \diff U$ maps measured in (a) forward and (b) backward scan direction with the same bias voltage $U=-200$~mV as in Fig.~\,\ref{fig:pic_2}(a,b). We again see areas of different $\diff I / \diff U$ contrast as a result of different magnetic domains in the image area. This means that the electron density a few \AA ngstr\"om above the surface (at the position of the tip) is substantially spin-polarized, despite the fact that the carbon atoms are expected to carry a very small magnetic moment.\cite{weser_2010,bertoni_2004,weser_2011} We conclude that we probe the magnetic structure of Ni(111) under the graphene layer. Evaluating the magnetic signal strengths for the $\diff I / \diff U$ maps in Fig.~\,\ref{fig:pic_4}(a,b), we find a magnetic asymmetry of $A(-200 \textnormal{mV})=4$~\%. We observe two types of boundaries between homogeneously magnetized areas: while the left one can clearly be identified as a domain wall with a width of $w=51\pm 6$~nm (see line profile and fit\cite{fit_with_angle,kubetzka_2003} in Fig.~\,\ref{fig:pic_4}(c)), the right one displays a non-continuous transition which appears at different lateral positions in forward and backward scan direction. This sharp transition is therefore not a domain wall, but instead it is an artifact resulting from magnetostatic interaction with the tip (see sketches in Fig.~\,\ref{fig:pic_4}(a,b)). Its lateral position depends on the scan direction (i.e. from left to right or vice versa) since the domain wall is pushed along the scan direction by the tip until it snaps back. This means that the tip used in this experiment exhibits a non-negligible stray field, most likely due to picking up magnetic material from the sample. Such an influence is frequently observed when using ferromagnetic tips to investigate soft magnetic materials both in SP-STM and MFM.\cite{wiesendanger_2009} Reasons for the two walls in the image area responding non-equivalently might be different magnetization directions or different wall types. The fact that the magnetic structure of Ni(111) can easily be changed by small external magnetic fields (cf.~Fig.~\,\ref{fig:pic_2}(b)) suggests that already small amounts of ferromagnetic material at the tip apex are sufficient to observe interactions with the domain structure.

\section{\label{sec:theory_method}Computational method}
To gain a detailed understanding of the observations made in our STM experiments we performed DFT calculations of the pure Ni(111) and the graphene/Ni(111) system using the projector augmented wave\cite{bloechl_paw} based Vienna-ab-initio-Simulation-Package (VASP).\cite{kresse_hafner,kresse_joubert} We employed the generalized gradient approximation (GGA)\cite{pbe} to the exchange correlation potential and accounted for van der Waals interactions in the calculations involving graphene in the framework of the DFT-D2 method.\cite{dftd2} We calculated the electronic properties of a clean Ni(111) slab and a Ni(111) slab coated with graphene on one side both containing 15 atomic layers of Ni and $\sim 18$~\AA~of vacuum between periodic images of the slabs. 

The triangular unit cell of the fcc Ni(111) surface and the triangular graphene unit cell exhibit a lattice mismatch of 1.8~\%. In our calculations, we used a unit cell with the experimental lattice constant of the Ni(111) surface $a=2.49$~\AA~\cite{khomyakov_2009} and the graphene layer deposited in the so-called \textit{top-fcc} arrangement (Fig.~\ref{fig:pic_3}(d)), which has been established as the energetically most favorable one in experimental and theoretical investigations.\cite{gamo_1997,fuentes_2008} All geometries considered in our calculations were optimized until all forces were below $0.01$~eV~\AA$^{-1}$. For obtaining the LDOS the Brillouin zone integrations were performed with the tetrahedron method\cite{bloechl} using $36\times36$ $\mathbf{k}$-point meshes.

We simulated high-resolution STM images by calculating the position dependent LDOS in the vacuum and integrating the energy window from -100 meV to 100 meV. 
To simulate STM spectra, we have calculated the LDOS inside so-called empty spheres placed at $3.6$~\AA~ above the surfaces in the vacuum region of the slab. The LDOS inside the empty spheres was calculated assuming an STM tip with an $s$-wave symmetric apex state and we carefully checked that the LDOS at lower/higher distances (between 2~\AA~ and 7~\AA) from the slab shows a smooth trend and yields qualitatively the same values. We considered spheres at four different lateral positions above the surfaces (see red dots in Fig.~\ref{fig:pic_3}(d)) to investigate lateral modulations in the STM spectra.

In STM experiments, there is an electric field between a tip and a sample, which modifies the shape of the tunneling barrier.\cite{olesen_1996} The exact shape of the tunneling barrier is unknown but can be approximated by a trapezoidal barrier in the most simple model. Besides any density of states effects, this leads to an energy and tip-height dependence of the $\diff I / \diff U$-signal according to\cite{rosink_2000} 
\begin{equation}
\frac{\diff I}{\diff U} \sim \exp\left(-\int\limits_0^s\diff z\left[\frac{8m}{\hbar^2}\left(\Phi+eU\frac{z}{s}-eU\right)\right]^{-\frac{1}{2}}\right),
\end{equation}
where $\Phi$ is the work function of the tip and the sample and $s$ is the distance between the tip and the sample.
Expanding the exponent to first order in $U$ we arrive at
\begin{equation}
\frac{\diff I}{\diff U}\sim c_0\exp\left(-\frac{eU}{E_0}\right),
\end{equation}
where $E_0$ is a constant depending on the materials of the tip and the sample as well as on their distance. 

To simulate STM spectra, we therefore use the vacuum LDOS from our DFT calculations (which accounts for the tunneling barrier due to the sample work function) and scale this vacuum LDOS by a factor of $\exp(-E/E_0)$ to account for electric field induced dependencies of the effective tunneling barrier height on the bias voltage. $E_0$ is treated as a fitting parameter. In Section \ref{sec:theory} we use $E_0=2$~eV, which leads to good agreement of our first-principles calculations with the experimental data.

\section{\label{sec:theory}Theory results}
\begin{figure}[t]
\includegraphics[width=1.0\columnwidth]{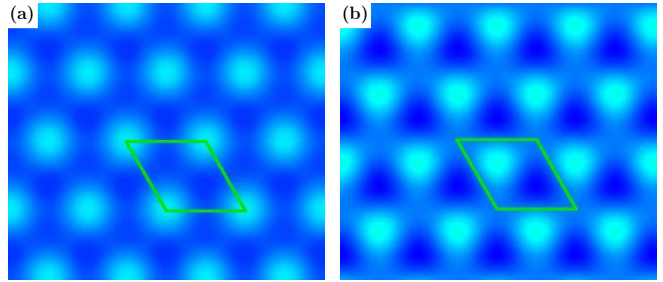}
\caption{\label{fig:stm_sim} {(color online) Simulated STM images calculated at a height of 3.6~\AA~over (a) clean Ni(111) and (b)~graphene/Ni(111). The orientation of the graphene lattice is the same as in the unit cell depicted in Fig.~\ref{fig:pic_3}(d): the C atoms visible in the image belong to the B sublattice.}}
\end{figure}

We start with the simulated STM images (Fig.~\ref{fig:stm_sim}) and compare them to the experiments (Figs.~\ref{fig:pic_1}(b) and \ref{fig:pic_3}(b)). Over the clean Ni(111) surface, we find a triangular lattice of protrusions visible as bright spots (Fig.~\ref{fig:stm_sim}(a)), which is in good agreement with the experimental STM image (Fig.~\ref{fig:pic_1}(b)). Our calculations show that the protrusions are centered above the atoms of the topmost Ni layer.

Our calculations as well as Refs.~\onlinecite{gamo_1997,khomyakov_2009} yield graphene adsorbing at a distance of $\sim2.10$~\AA~above the Ni surface with the graphene layer being almost flat, the height difference between the carbon sublattices (a structural corrugation) is about $0.5$~pm. However, due to an electronic effect, the simulated STM images of graphene/Ni(111) also show a triangular lattice structure (Fig.~\ref{fig:stm_sim}(b)) again in good agreement with the experiment (Fig.~\ref{fig:pic_3}(b)). In the convention of Fig.~\ref{fig:pic_3}(d), our calculations yield the highest vacuum LDOS above carbon atoms of the B sublattice, i.e. those carbon atoms not located above a Ni atom from the topmost Ni(111) layer. This feature is stable with the simulated tip height and bias voltage. We thus conclude that the highest protrusions in the STM images of graphene on Ni(111) correspond to carbon atoms in sublattice B.

In the STM experiments of graphene on Ni(111), the graphene sublattices A and B exhibit an apparent height difference on the order of $10$~pm. Since the structural corrugation is only about 0.5 pm, this enhanced corrugation in the experiments must have an electronic but \textit{not} a structural origin. The electronic states responsible for this corrugation will be discussed together with the calculated STM spectra below.
\begin{figure}[t]
\includegraphics[width=1.0\columnwidth]{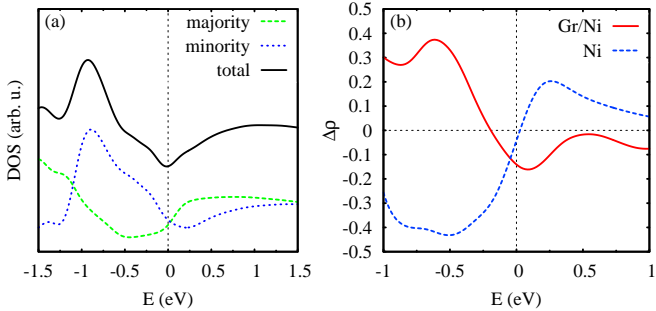}
\caption{\label{fig:spincontrast} {(color online) (a) Simulated STS spectrum at $3.6$~\AA~ above the clean Ni(111) surface. (b) Spin contrast $\Delta\rho$ (see text) at $3.6$~\AA~over the pristine Ni(111) surface and the graphene-coated Ni(111) surface respectively. }}
\end{figure}

\begin{figure*}[t]
\includegraphics[width=1\linewidth]{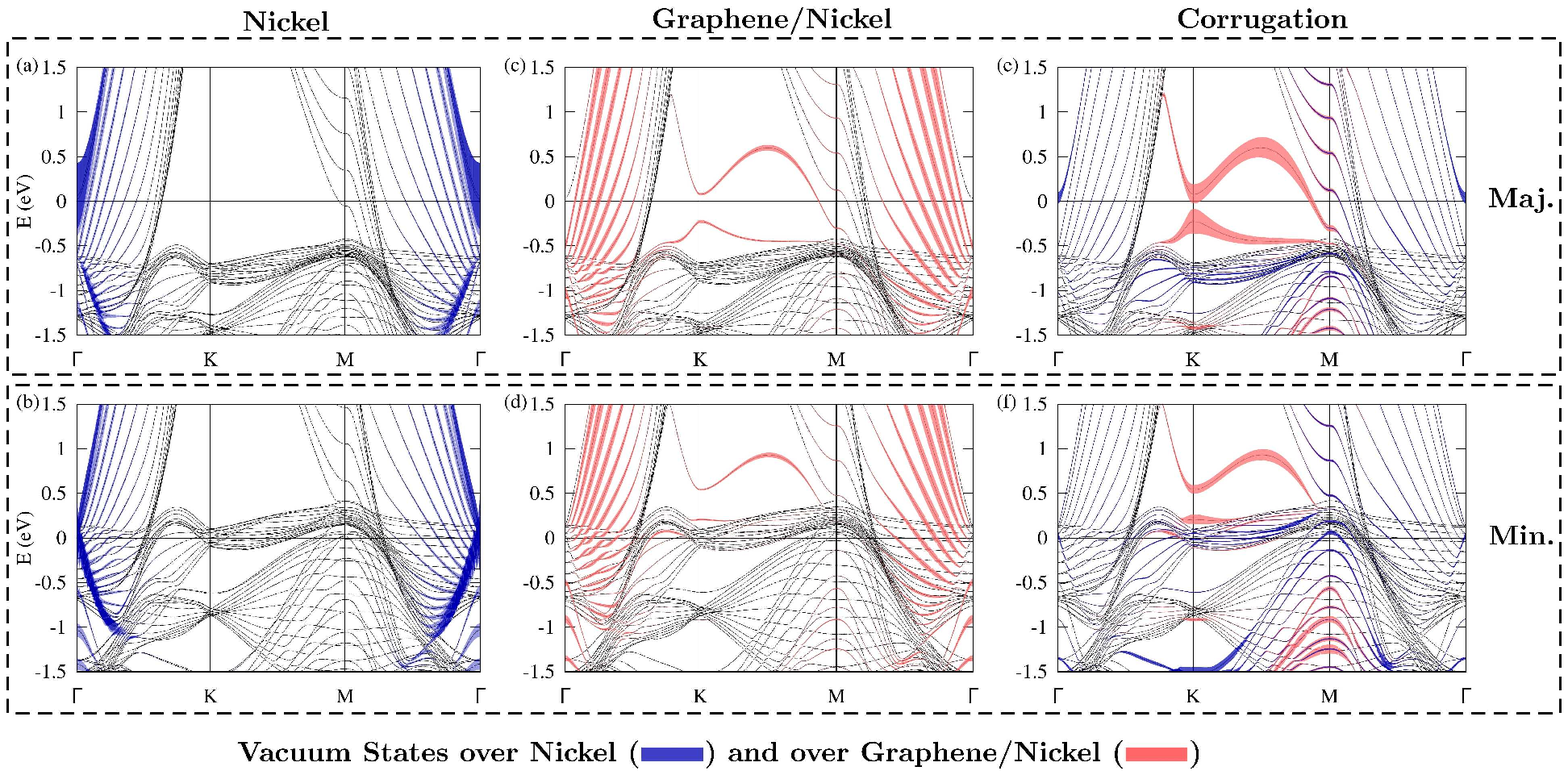}
\caption{\label{fig:bands} {(color online) Band structure of a clean 15-layer slab of Ni(111) for majority (a) and minority spin (b) components. The same slab, coated with graphene on one side, is shown in (c) for the majority and in (d) for the minority case. The projections of the bands onto empty spheres in the vacuum at 3.6~\AA, averaged over their lateral position (red dots in Fig.~\ref{fig:pic_3}(d)), are shown as fat bands. Here, the vacuum projections above the Ni are indicated by blue color, over graphene by red color. In (e) and (f) the contributions of each band to the corrugations measured in STM are visualized as fat bands (see Eq.~(\ref{eq:corrug})) in the same color code. For visualization purpose, the corrugation above Ni has been enhanced by a factor of 4.}}
\end{figure*}

In the experiment we obtained energy resolved STM spectra and spin-resolved differential conductance ($\diff I / \diff U$) maps, which we now compare to spectra from our DFT calculations. The calculated STS spectrum for the pristine Ni(111) surface is shown in Fig.~\ref{fig:spincontrast}(a). It exhibits broad maxima in the energy range between -1~eV and -0.5~eV as well as above $E_\textnormal{F}$ at energies $\gtrsim 0.5$~eV. We find that the Ni spectrum does not change with the lateral position of the tip. A comparison to Fig.~\ref{fig:pic_1}(d) shows that the calculated spectrum is in qualitative agreement with the experimental results.

Our calculations show that the contribution to the STS signal arising from majority (spin-up) and minority spin states (spin-down) differs clearly, which is well in line with spin contrast being achievable in our SP-STM experiments. To simulate the magnetic contrast provided by the SP-STM we calculated the spin contrast $\Delta\rho=(\rho_{\uparrow}-\rho_{\downarrow})/(\rho_{\uparrow}+\rho_{\downarrow})$ from the spin polarized density of states, $\rho_{\uparrow\downarrow}$. The spin contrast, averaged over the four different spheres in the vacuum (Fig.~\ref{fig:pic_3}(d)), for the clean Ni(111) surface and the graphene-coated surface is shown in Fig.~\ref{fig:spincontrast}(b). In agreement with the experiment we find a pronounced spin contrast over clean Ni as well as over the graphene-coated surface. Thus, the spin polarization in the vacuum above the surface is not suppressed by the graphene coating and the ferromagnetic domain structure remains visible in SP-STM. Interestingly, the sign of the spin polarization in the vacuum LDOS is reversed for graphene-coated Ni as opposed to pristine Ni at energies below -0.23~eV as well as above $E_\textnormal{F}$. This spin contrast change may be accessed in future experiments, where one would have to perform an SP-STM measurement on a partially graphene-coated and partially clean Ni(111) sample.

We now aim at identifying which states of the Ni(111) and the graphene/Ni(111) systems give major contributions to the tunnel current in the STM experiments and thus understand the physics behind the calculated and observed STM images, spectra and spin contrasts. To this end, we calculated the band structure and analyzed the corresponding wave functions of the clean and the graphene-coated Ni surface. The band structure of a clean Ni slab is shown in Fig.~\ref{fig:bands} for majority (a) and minority (b) spin states. In this figure we use a so-called "fat band analysis"\cite{andersen_1995} where the displayed thickness of a band represents the strength of the property of interest. Here the wave function, $\ket{\Psi_{n,k}}$, belonging to a given band $n$ at a given $k$-point is projected onto an $s$ orbital, $\ket{L}$ , localized inside an empty sphere at $3.6$~\AA~above the respective surface. The overlap $|\hpr{\Psi_{n,k}}{L}|^2$ is then depicted as the thickness of the corresponding band. 

For clean Ni(111), the dominant contributions to the vacuum LDOS (blue) above $E_\textnormal{F}$ arise from upward dispersing bands, which have their energy minima at the $\Gamma$ point at energies of 10~meV (majority spin electrons) and 140~meV (minority spin electrons). These states can be characterized as surface states or surface resonances. The upward dispersing feature with a minimum at the $\Gamma$ point for majority spin is the well-known Shockley surface state with mixed Ni $p_z$ and $d_{3z^2-r^2}$ character at $\Gamma$ and $d_{xz,yz}$ admixtures away from $\Gamma$. \cite{magaud_2004,ohwaki_2006}
A similar feature for the minority spin was identified as a resonance of mixed $p_z$ and $d_{xz,yz}$ character.\cite{magaud_2004,ohwaki_2006}

The downward dispersing feature starting at -0.6~eV below $E_\textnormal{F}$ for the majority states and dispersing along the $\Gamma\to\mathrm{K}$ and $\Gamma\to\mathrm{M}$ directions in the Brillouin zone was identified as a surface resonance.\cite{braun_2002} This specific resonance derives mainly from Ni $d_{xz,yz}$ states with a small admixture of Ni $p_z$ states away from the $\Gamma$ point. At the $\Gamma$ point itself, however, it shows no spectral weight.  
For the minority spin component a similar state begins slightly above $E_\textnormal{F}$ and disperses down below -1~eV along the $\Gamma\to K$ and $\Gamma\to M$ directions. This feature, showing pronounced weight in the vacuum, was identified as a surface resonance with the same orbital characteristics as the Shockley state described above.\cite{ohwaki_2006,braun_2002}
This state is responsible for the minority spin polarization dominating in the energy region from $\sim -1$~eV up to $\sim 0.1$~eV seen in the spin contrast in Fig.~\ref{fig:spincontrast}(b). The STS spectrum shown in Fig.~\ref{fig:spincontrast}(a) also becomes clear now: The features of the spin-resolved spectra can be attributed to surface electronic features of Ni. The broad peak below $E_\textnormal{F}$ stems mostly from the downward dispersing feature of the minority spin states in Fig.~\ref{fig:bands}(b), whereas the unoccupied spectrum is dominated by the Shockley state of both spin character.

Over the graphene-coated surface the situation changes qualitatively. Fig.~\ref{fig:bands}(c,d) show the fat band analysis for the Ni slab coated on one side with graphene. Over graphene, we find dominant contributions to the spectral weight in the vacuum (red fat bands) arising from free electron like states as can be seen along $\Gamma\to K$ and $\Gamma\to M$ directions. The graphene $p_z$-derived bands between the $K$ and $M$ points show some smaller weight in the vacuum as well. The pronounced surface states and resonances seen over the Ni surface cannot be seen over graphene, even at very low heights over the surface. The band derived from the Shockley state vanishes explicitly over the graphene-coated slab surface. Only a band from the uncoated surface of the slab remains. Thus, graphene quenches the surface resonances and surface states. Since the surface states are mainly responsible for the sign of the spin contrast in the vacuum over the Ni surface the quenching of these states leads to the predicted reversal of the sign of the contrast (cf. Fig.~\ref{fig:spincontrast}(b)). 

We note that this reversal of the spin contrast in the vacuum LDOS above graphene does not mean that the Ni magnetization is reversed beneath graphene. We find in agreement with earlier studies \cite{bertoni_2004} that the magnetic moment in the Ni interface layer to graphene is about 20~\% smaller (0.51$~\mu_\textnormal{B}$) compared to the bulk value of 0.65$~\mu_\textnormal{B}$. Additionally, a small spin magnetic moment of -0.02$~\mu_\textnormal{B}$ for $C_A$ and 0.03$~\mu_\textnormal{B}$ for $C_B$ atoms respectively is induced in the graphene layer.\cite{bertoni_2004,weser_2011}
 
We finally address the question why the apparent height corrugations measured in STM (see Section \ref{sec:stm}) are larger in the graphene/Ni system than on the bare Ni(111) surface. To this end, we compared the vacuum amplitudes of the states of clean Ni(111) and graphene-coated Ni systems at different lateral positions within the unit cell (red dots in Fig.~\ref{fig:pic_3}(d)). The variation of the state amplitude with the lateral position is visualized in Fig.~\ref{fig:bands}(e,f): for each band $\ket{\Psi_{n,k}}$, the thickness $d$ of the colored curves (blue: Ni, red: graphene) corresponds to the standard deviation of the vacuum projections $|\hpr{\Psi_{n,k}}{L_r}|^2$ with lateral sphere position $r$: 
\begin{equation}
d\sim\sqrt{\sum_r\left(|\hpr{\Psi_{n,k}}{L_r}|^2-\overline{|\hpr{\Psi_{n,k}}{L_r}|^2}\right)^2}/\rho_{\uparrow\downarrow}(E).
\label{eq:corrug}
\end{equation}
The line thicknesses are normalized to the (Brillouin zone integrated) local density of states $\rho_{\uparrow\downarrow}(E)$ at the respective energies.

The variation of the wave function amplitudes with the lateral sphere position is clearly higher over graphene than above clean Ni, which can be seen in Fig.~\ref{fig:bands}(e,f). It is visible that the corrugations measured above graphene on Ni mainly arise from graphene derived $p_z$ states. Similar to the case of graphite,\cite{tomanek_1987} the peculiar symmetry of these states induces an asymmetry in STM images. For small bias voltages, hence close to $E_\textnormal{F}$, the upper graphene $p_z$-derived band crossing $E_\textnormal{F}$ near the M point contributes strongly to the corrugation in the majority spin channel. The lower $p_z$ band and small contributions from Ni $d$ bands induce the corrugation in the minority spin case. Above clean Ni(111) our STM experiments measure smaller corrugations, which mainly originate from slight lateral variations in the surface states near the center of the Brillouin zone. It becomes clear that the corrugations above the graphene sheet are of electronic origin and are also brought about by the quenching of the Ni surface states as can be seen in our data.

\section{\label{sec:kerr}Kerr-Microscopy results}

\begin{figure}[t]
\includegraphics[width=1\columnwidth]{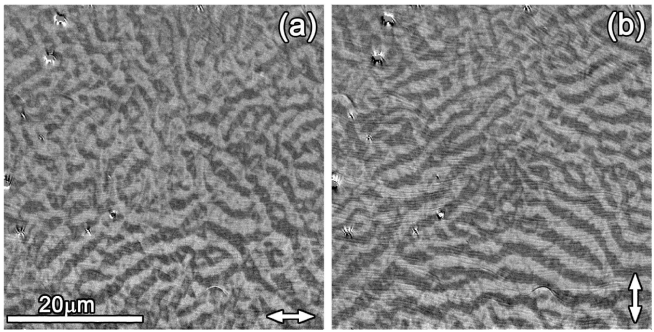}
\caption{\label{fig:kerr} {Kerr microscopy images of the magnetic structure of the Ni(111) single crystal.
The gray-scale amplitude of the images is proportional to the in-plane magnetization component which is indicated by a double arrow in the lower right corner.}}
\end{figure}

\begin{figure*}[t]
\includegraphics[width=2\columnwidth]{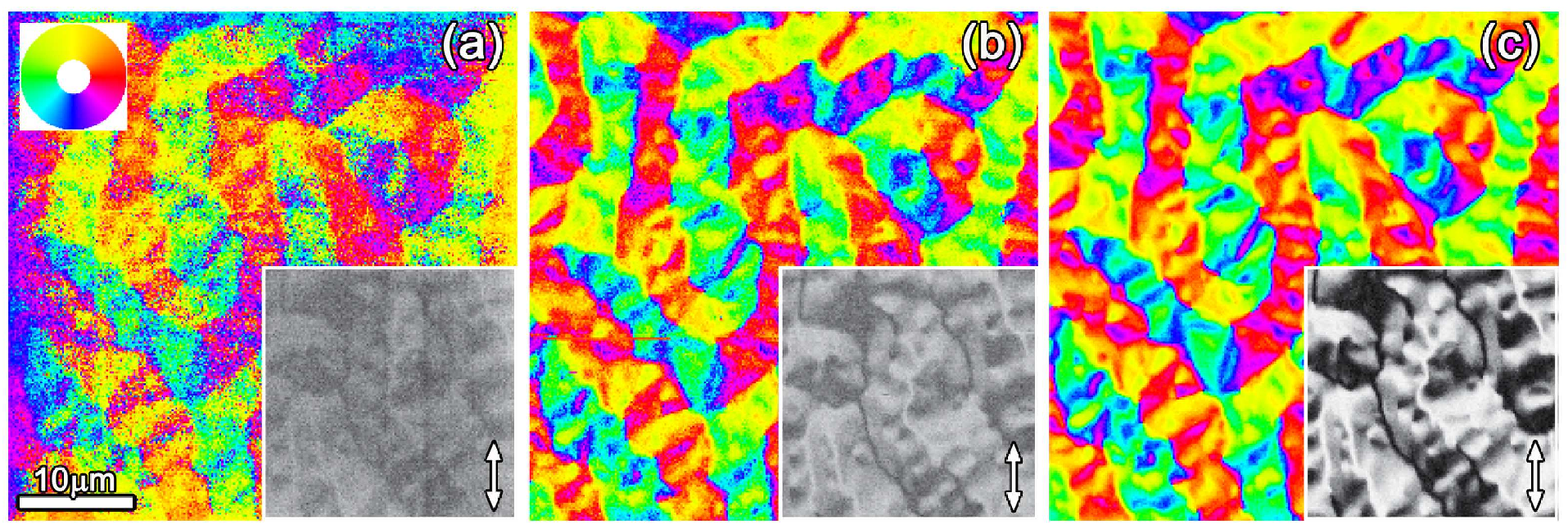}
\caption{\label{fig:sempa} {(color online) SEMPA images of the magnetic structure of the Ni(111) surface at the same position, but with different surface preparations. 
Encoded in color is the direction of the in-plane magnetization, as defined by the 360$^\circ$ color wheel. In (a) the surface is covered with graphene. (b) gives the
signal from the clean Ni surface following sputter-cleaning. For contrast enhancement, in (c) a thin iron layer has been deposited. To emphasize the changes in
contrast, the lower right part of each image gives the y-component of the magnetization in a gray-scale representation.}}
\end{figure*}

To identify the large-scale magnetic domain structure we studied the domain pattern of Ni(111) by means of Kerr microscopy. An in-plane measurement shows a recurring magnetic pattern that varies slightly for different demagnetization cycles while we found no indication for an out-of-plane component. In Fig.~\ref{fig:kerr} the horizontal (a) as well as the perpendicular component (b) of the in-plane magnetization is shown. The in-plane magnetic pattern is characterized by magnetic structures of two length scales: we observe a stripe domain pattern with a stripe width in the range of 3~$\mu$m to 6~$\mu$m which varies the orientation in different sample areas. Inside each of these stripes one can see a wavy pattern indicative of a magnetic fine structure on a smaller length scale. A real-time observation of the Ni(111) single crystal during the application of an external magnetic in-plane field shows that at about 13~mT the domain walls begin to move. A field of 140~mT is sufficient to create a single domain state.

\section{\label{sec:sempa}SEMPA results}

\begin{figure}[t]
\includegraphics[width=1\columnwidth]{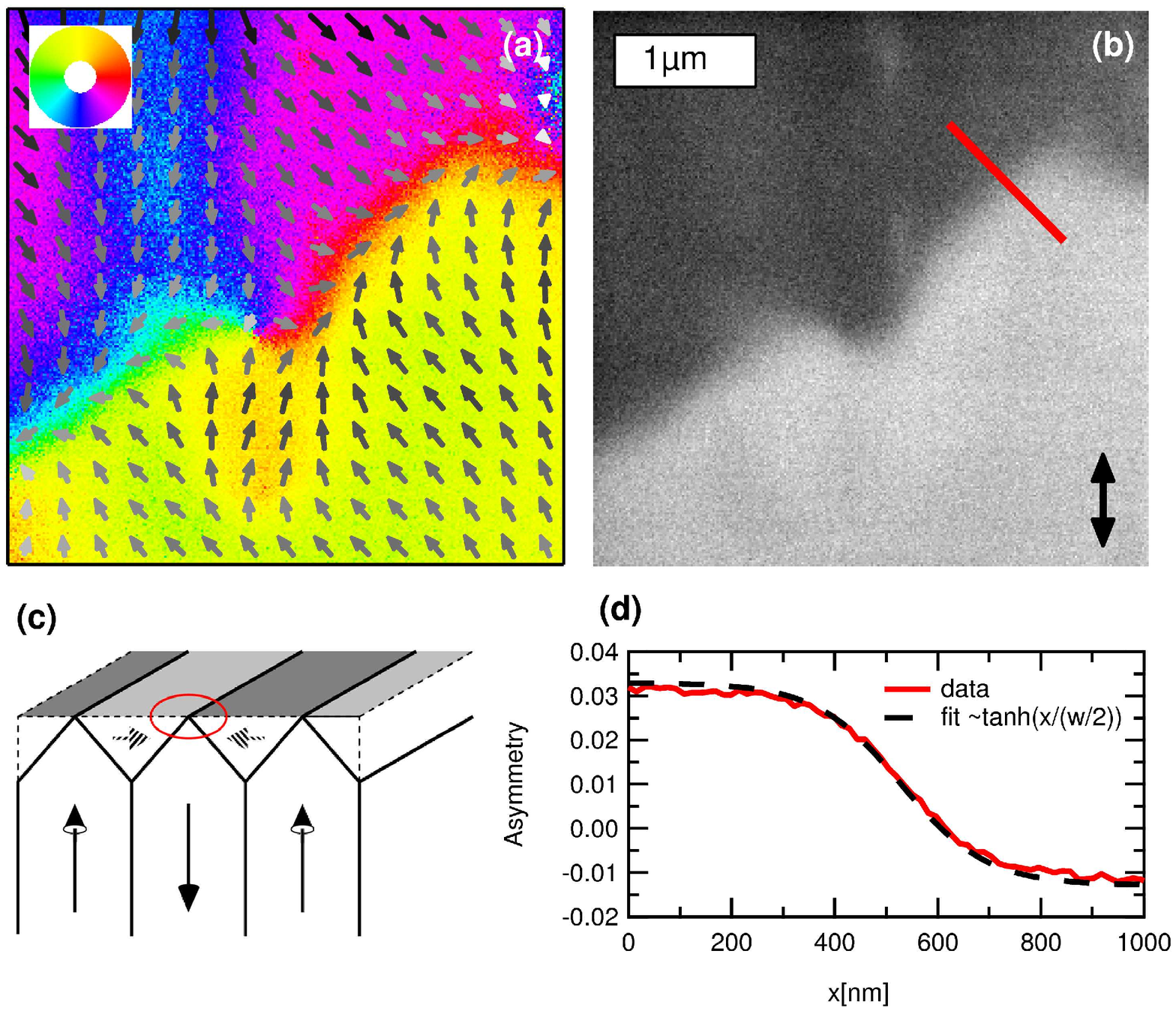}
\caption{\label{fig:sempaWall} {(color online) High resolution SEMPA images of the N\'eel-like walls at the Ni(111) surface. In (a) the in-plane distribution of the
magnetization of the Ni(111) surface is shown. Encoded in color is the direction of the in-plane magnetization, as defined by the 360$^\circ$ color
wheel. In addition the direction is illustrated by the arrows. In (b) the y-component of the magnetization is shown in a gray-scale representation. The striped line is indicating the path of the wall profile shown in (d). (c) Sketch of a section across a V-line on the surface, illustrating the corresponding volume domain structure. The striped arrows indicate the branched structure of the corresponding domains. (d) The wall-profile is fitted by Eq.~(\ref{eq:one}) This fit yields a wall width $w$ = 330~$\pm$~6~nm. }}
\end{figure}

To get a more detailed picture of the surface magnetic domain structure we performed high-resolution SEMPA measurements. The SEMPA image in Fig.~\ref{fig:sempa}(a) shows a characteristic section of the magnetization pattern of the graphene-covered Ni(111) single-crystal surface, after transfer under ambient conditions. Although no further treatment of the surface has been performed, we observe a meaningful magnetic contrast, i.e. a polarization asymmetry of 1.2~\%. The observation of magnetic contrast in SEMPA without in situ cleaning of the sample is most unusual, because of the sub-nm surface sensitivity of secondary-electron spin polarization. 

Therefore we must deduce that the Ni(111) surface is effectively passivated by the graphene layer, which is in accordance with recent spectroscopy data.\cite{dedkov_2008} To check for any graphene-induced change of signal and/or domain pattern Figure~\ref{fig:sempa}(b) shows the same area of the sample as in (a) after argon ion sputtering. While the domain structure is unchanged, the clean Ni surface exhibits a much stronger magnetic contrast corresponding to 2.4~\% asymmetry. The sign of the secondary-electron spin polarization is preserved, which indicates, that in the SE cascade process the graphene does not cause a polarization inversion. This should not be confused with the calculation given in Fig.~\ref{fig:spincontrast}, as SEMPA detects free electrons in vacuum, which have an energy $>$ 5~eV with respect to the Fermi level. The image quality can be improved by depositing a small amount of iron ($\approx$ 4~AL) onto the surface. Due to its higher saturation magnetization and the sub-nm surface sensitivity of SEMPA the iron acts as polarizer for the secondary electrons. The result can be seen in Fig.~\,\ref{fig:sempa}(c) and again the domain pattern is identical to (a) and (b), while the contrast is enhanced to give a polarization asymmetry of 6.7~\%. These differences of the magnetic contrast and thus signal-to-noise ratio are highlighted by the gray-scale images in the bottom right corner, which display the experimental results on the same scale.

The SEMPA measurements confirm the results from Kerr microscopy and provide images of the surface magnetic domain pattern at higher resolution: it 
consists of a larger length-scale stripe pattern with a width from 3~$\mu$m to 6~$\mu$m. However, these larger stripes are not single 
domains but instead the magnetization within the stripes is more or less regularly modulated by a second type of stripes, on a smaller length-scale from 1~$\mu$m to 3~$\mu$m. From one of these small stripes to the next, the magnetization changes by approximately 60 degrees, which results in a net magnetization of the large stripes. The 
transitions between the smaller stripes are very broad yielding a rather wavy pattern without sharp domain walls. In contrast, the large stripes are separated by sharp  
N\'eel-like walls, which are of the 180$^\circ$ type. Figure~\ref{fig:sempaWall}(a) shows a higher magnification of the domain structure around such a head-to-head domain wall. The \textit{y}-component of the magnetization is shown in (b). The striped line indicates the position of the wall profile plotted in Fig.~\ref{fig:sempaWall}(d). It can be described by Eq.~(\ref{eq:one}), and a fit yields a domain wall width $w$ = 330$~\pm$~6~nm. The observation of much narrower domain walls by SP-STM does not contradict the SEMPA results: in first order approximation the width of a N\'eel wall is proportional to $K_{1}^{-1/2}$, where $K_{1}$ is the first order magneto crystalline anisotropy constant. From Refs.~\onlinecite{pussei_1957,franse_1968} it is known that $K_{1}$ is strongly temperature dependent between 300~K and 4~K: -0.0045~MJ/m$^{3}$ at RT compared to -0.12~MJ/m$^{3}$ at 4~K. Therefore, a reduction of wall width with temperature is expected. Using the width from the SEMPA measurements as starting point, we estimate a wall width at 4~K of approximately 70~nm, in reasonable agreement with the wall width found using SP-STM.

To reveal the origin of the observed surface magnetic pattern it is crucial to understand why the N\'eel walls have a head-to-head or tail-to-tail configuration 
as seen in the SEMPA images. If this pattern persisted into the volume, it would imply a huge amount of dipolar energy. Instead, this pattern is indicative 
of a so-called V-line structure,\cite{hubert_1998} where two volume domain walls with different orientations merge at the surface into a single line. 
In a cross-section perpendicular to the line, this structure appears like a V where the magnetization of the center domain collects all the flux that 
originates from the oppositely magnetized side domains (see Fig.~\ref{fig:sempaWall}(c)). So the surface pattern is flux compensated in the volume to reduce the dipolar energy. As the magnetic surface structure of the V-lines imaged in Fig.~\ref{fig:sempaWall} shows only an in-plane magnetization we can interpret it as the N\'eel cap of a V-line, in analogy to the well-understood N\'eel cap of a Bloch wall e.g.~in Fe(001).\cite{oepen_1989} Indeed, our measurements give no indication of an out-of-plane component of the magnetization in the domains or the domain walls, in agreement with the Kerr-microscopy measurement. Thus we conclude that the magnetization at the surface is entirely in-plane. This may be surprising, as none of the magnetically easy $\left\langle 111 \right\rangle$-directions of the Ni crystal lies within the (111) surface of the sample. This finding can only be explained as a consequence of a reduction of stray field energy at the expense of local anisotropy within a certain depth of subsurface volume. Estimating the upper limit of the shape anisotropy of the crystal with $\frac{\mu_{0}}{2} M_{s}^{2}$ and comparing this to the magneto-crystalline anisotropy constant in first order $K_{1}$ at RT we obtain a ratio of 34:1. This might explain the in-plane orientation of the observed magnetic pattern.

The above-described features of the magnetic pattern of the surface of the Ni(111) crystal can be understood qualitatively using the quasi-domain branching approach for 
large crystals with strongly misoriented surfaces given by Hubert and Sch\"afer. \cite{hubert_1998} For a Ni platelet with (111) surfaces the model assumes 
180$^\circ$-oriented base domains in the volume, which reduce the magnetostrictive energy of the crystal. The lateral extent of these domains would be responsible for the length scale of the larger stripes we find on the surface. In the mentioned quasi-domain branching concept, quasi-domains with a net magnetization parallel to the surface are introduced, which close the flux of the basis volume domains. They are composed of alternating domains oriented along the easy directions. Each of these first-level branching domains acts as basis domain for second-level branching and thus forms its own closure quasi-domain at the surface to further reduce the stray-field energy. In the model, the energy gain by branching of the closure domains in comparison to the amount of domain wall energy needed for further branching determines the branching depth that is finally observed. In our measurements, we observe a two-level branching only, where at least the surface of the second level domains is already fully in-plane oriented. This is in contrast to the model of Hubert, where quasi domains along the out-of-plane canted easy axes are expected even on the final level of branching.

\section{\label{sec:summary}Summary}
Finally, we would like to summarize the main aspects of this study.~(i) We probed the differential conductance of bare Ni(111) by STM and in comparison to our DFT calculations attributed the observed features to a minority spin surface resonance below and the Shockley state of both spin character above $E_\textnormal{F}$.~(ii) For graphene/Ni(111), STM images showed a triangular lattice of the atomic structure and an enhanced corrugation compared to pure Ni(111). DFT showed that both properties are a purely electronic effect originating from  graphene $p_z$ states around $E_\textnormal{F}$ and the quenching of Ni surface states.~(iii) Kerr microscopy and SEMPA measurements revealed an entirely in-plane magnetic pattern of Ni(111) which stems from a two-level branching of the domain structure.~(iv) We could easily move domain walls in low external magnetic fields or by the stray field of a magnetic tip. A single in-plane domain state was observed at RT in an in-plane magnetic field of 140 mT.~(v) The reactive properties of Ni(111) are passivated by the graphene layer well enough for magnetic imaging after transfer through air. The magnetic structure of graphene-coated Ni is unchanged compared to that of bare Ni(111).~(vi) As a result of surface state quenching, our DFT calculations predict an inversion of spin-polarization above the graphene layer with respect to the pristine Ni(111) surface.

\section{\label{sec:ack}Acknowledgements}
Financial support from the SFB 668 and grant WI1277/25 of the Deutsche Forschungsgemeinschaft and from the Hamburg Cluster of Excellence NANOSPINTRONICS are gratefully acknowledged.

\bibliography{gni111}

\begin{thebibliography}{45}
\bibitem [{sta()}]{standard_fit}%
  \BibitemOpen
  \href@noop {} {\bibinfo  {journal} {We used a fit for a 180$^\circ$ Bloch
  wall for the only purpose of estimating the domain wall width}\ }\BibitemShut
  {NoStop}%
\bibitem [{fit()}]{fit_with_angle}%
  \BibitemOpen
  \href@noop {} {\bibinfo  {journal} {There is a fit of a general wall profile
  measured by SP-STM: cos(arcsin(tanh((x)/(w/2))+$\theta$), where $\theta$ - an
  angle between a tip and a domain wall magnetizations. As a result of fitting,
  the angle $\theta$ was extracted to be 72$^\circ$. In case when $\theta$ = 0,
  this fit can be simplified into a fit of a standard wall profile, thus, is
  described by Eq. (1)}\ }\BibitemShut {NoStop}%
\end{thebibliography}

\end{document}